\documentclass[12pt]{article}
\usepackage{graphicx}

\newsavebox{\sboxpubnumber}
\newsavebox{\sboxpubdate}

\newcommand{\Title}[1]{\begin{center} {\Large #1 } \end{center}}
\newcommand{\Author}[1]{\begin{center}{ \sc #1} \end{center}}
\newcommand{\Address}[1]{\begin{center}{ \it #1} \end{center}}

\newenvironment{Abstract}{\begin{quotation}  }{\end{quotation}}

\begin{document}

\begin{titlepage}

\vfill
\Title{A nonextensive entropy approach to kappa-distributions}
\vfill
\Author{Manfred P. Leubner}
\Address{Institute for Theoretical Physics, University of Innsbruck\\
         A-6020 Innsbruck, Austria}
\vfill
\vfill
\begin{Abstract}
Most astrophysical plasmas are observed to have velocity distribution
functions exhibiting non-Maxwellian suprathermal tails. The high energy
particle populations are accurately represented by the family of
kappa-distributions where the use of these distributions has been 
unjustly criticized because of a perceived lack of theoretical
justification. We show that distributions very close to kappa-distributions
are a consequence of the generalized entropy favored by nonextensive statistics,
which provides the missing link for power-law models of non-thermal features
from fundamental physics. With regard to the the physical basis supplied
by the Tsallis nonextensive entropy formalism we propose that this slightly
modified functional form, qualitatively similar to the traditional
kappa-distribution, be used in fitting particle spectra in the future.
\end{Abstract}

\vfill
\end{titlepage}

\def\thefootnote{\fnsymbol{footnote}}
\setcounter{footnote}{0}

\section{Introduction}

A variety of space observations indicate clearly the ubiquitous presence of
suprathermal particle populations in astrophysical plasma environments
\cite{Mendis94}. The family of kappa velocity space distributions,
introduced first by Vasyliunas \cite{Vasyliunas68}, is recognized to be highly appropriate
for modeling specific electron and ion components of different plasma states.

Numerous magnetospheric interaction processes and
instabilities were studied successfully within the concept of
kappa-distributions \cite{Xue96,Chaston97} ranging from plasma
sheet ion and electron spectra \cite{Christon91} to a combined
electron-proton-hydrogen atom aurora \cite{Decker95}. The saturation
of ring current particles towards a kappa-distribution was
studied by Lui and Rostoker \cite{Lui95} and accurate fittings of observed
electron flux spectra were performed by a power law at high energies
\cite{Janhunen98}. Furthermore, it was demonstrated by Leubner \cite{Leubner00a}
that the Jovian banded whistler mode emission can be interpreted within a
kappa-distribution approach and that mirror instability thresholds are
drastically reduced in suprathermal space plasmas \cite{Leubner00b,Leubner01}.
Energetic tail distributions play a key role in coronal plasma dynamics
\cite{Maksimovic97} and high resolution plasma observations near 1 AU
confirm that even the distribution function of heavy solar wind ions
are well fitted by a kappa-distribution \cite{Collier96}.

The generation of velocity space distributions exhibiting pronounced
energetic tails was frequently interpreted as a consequence of several
different acceleration mechanisms. Those include besides DC parallel
electric fields or field-aligned potential drops in reconnection regions
also wave-particle interaction due to kinetic Alfv\'{e}n wave
turbulence \cite{Hui92,Leubner00c} and cyclotron interactions
\cite{Gomberoff95}. Clear evidence of the importance of suprathermal
particle populations in space plasmas have motivated the development of
a modified plasma dispersion function for kappa-distributions \cite{Summers91,Mace95}.

The use of the family of kappa-distributions to model the observed non-thermal
features of electron and ion structures was frequently criticized since a
profound derivation in view of fundamental physics was not available. A
classical analysis addressed to this problem was performed by Hasegawa \cite{Hasegawa85}
demonstrating that kappa-distributions turn out as consequence of the presence of
suprathermal radiation fields in plasmas and Collier \cite{Collier93} considers the generation
of kappa-like distributions using velocity space L\'{e}vy flights. Furthermore, a
justification for the formation of power-law distributions in space plasmas
due to electron acceleration by whistler mode waves was proposed by Ma and
Summers \cite{Ma98,Ma99a,Ma99b} and a kinetic theory was developed showing that
kappa-like velocity space distributions are a particular thermodynamic equilibrium
state \cite{Treumann99}. Here we demonstrate that the family of kappa-distributions
results as consequence of the entropy generalization in nonextensive statistics,
providing thus the missing link for the use of kappa-distributions from fundamental
physics.

\section{Theory}

A generalization of the Boltzmann-Gibbs-Shannon entropy formula for statistical equilibrium
was recognized to be required for systems subject to spatial or temporal long-range
interactions making their behavior nonextensive. Any extensive formalism fails
whenever a physical system includes long-range forces or long-range memory. In
particular, this situation is usually found in astrophysical environments and plasma
physics where, for example, the range of interactions is comparable to the size of
the system considered. A generalized entropy is required to possess the usual
properties of positivity, equiprobability, concavity and irreversibility but
suitably extending the standard additivity to nonextensivity.   

In view of the difficulties arising in this conjunction within the Boltzmann-Gibbs
standard statistical mechanics motivated Tsallis \cite{Tsallis88,Tsallis95} to introduce a
thermo-statistical theory based on the generalized entropy of the form

\begin{equation}
S_{q}=k_{B}\frac{1-{\sum }p_{i}^{q}}{q-1}  \label{1}
\end{equation}

where $p_{i}$ is the probability of the $i^{th}$ microstate, $k_{B}$
is Boltzmann's constant and $q$ is a parameter quantifying the degree of
nonextensivity and is commonly referred to as the entropic index. A crucial
property of this entropy is the pseudoadditivity such that

\begin{equation}
S_{q}(A+B)=S_{q}(A)+S_{q}(B)+(1-q)S_{q}(A)S_{q}(B)  \label{2}
\end{equation}

for given subsystems $A$ and $B$ in the sense of factorizability of the
microstate probabilities. For $q\rightarrow 1$ the standard
Boltzmann-Gibbs-Shannon extensive entropy is recovered as

\begin{equation}
S_{q}=-k_{B}{\sum p_{i}}\ln p_{i}  \label{3}
\end{equation}

Applying the transformation

\begin{equation}
\frac{1}{q-1}=-\kappa  \label{4}
\end{equation}

to equation (\ref{1}) and restricting to values $-1<q\leq 1$ yields the
generalized entropy of the form

\begin{equation}
S_{\kappa }=\kappa k_{B}({\sum }p_{i}^{1-1/\kappa }-1) \label{5}
\end{equation}

for $\frac{1}{2}<\kappa \leq \infty$. With regard to Silva \cite{Silva98}
one finds the corresponding one dimensional equilibrium velocity space
distribution in kappa notation as

\begin{equation}
f(v)=A_{\kappa }\left[ 1+\frac{1}{\kappa }\frac{v^{2}}{v_{th}^{2}}\right]
^{-\kappa }  \label{6}
\end{equation}

where the normalization constant reads

\begin{equation}
A_{\kappa }=\frac{N}{v_{th}}\frac{1}{\sqrt{\kappa }}\frac{\Gamma \left[
\kappa \right] }{\Gamma \left[ \kappa -1/2\right]} \label{7}
\end{equation}

Here $N$ denotes the particle density, $v_{th}=\sqrt{2k_{B}T/m}$ is the thermal
velocity where $T$ and $m$ are the temperature and the mass, respectively, of the species
considered. The case $\kappa =\infty$ corresponds to $q=1$ wherefrom the
Maxwell equilibrium distribution is recovered. Figure 1a illuminates schematically
the non-thermal behavior of the distribution function (6) for some values of the
spectral index kappa.

Consistently, the one-dimensional equilibrium velocity space distribution in the
q-nonextensive framework is written as

\begin{equation}
f(v)=A_{q}\left[ 1-(q-1)\frac{v^{2}}{v_{th}^{2}}\right] ^{1/(q-1)}  \label{8}
\end{equation}

where

\begin{equation}
A_{q}=\frac{N}{v_{th}}\sqrt{1-q}\frac{\Gamma \left[ 1/(1-q)\right] }{\Gamma %
\left[ 1/(1-q)-1/2\right]} \label{9}
\end{equation}

for $-1<q\leq 1$. In analogy, the isotropic three dimensional velocity space distribution is
found as

\begin{equation}
f({\mathbf {v}})=B_{\kappa }\left[ 1+\frac{1}{\kappa }\frac{v^{2}}{v_{th}^{2}}%
\right] ^{-\kappa }  \label{10}
\end{equation}

identical to the distribution function for a plasma in a suprathermal
radiation field discussed by Hasegewa (1985) where the normalization constant reads

\begin{equation}
B_{\kappa }=\frac{N}{\pi ^{3/2}v_{th}^{3}}\frac{1}{\kappa ^{3/2}}\frac{%
\Gamma \left[ \kappa \right] }{\Gamma \left[ \kappa -3/2\right]} \label{11}
\end{equation}

for $\frac{3}{2}<\kappa \leq \infty$. Equation (\ref{6}) and (\ref{10}) denote a reduced
form of the standard kappa-distribution used for astrophysical applications and
written conventionally as

\begin{equation}
f({\mathbf {v}})=A_{\kappa }\left[ 1+\frac{1}{\kappa }\frac{v^{2}}{v_{th}^{2}}%
\right] ^{-(\kappa +1)}  \label{12}
\end{equation}

with the normalization constant

\begin{equation}
A_{\kappa }=\frac{N}{\pi ^{3/2}v_{th}^{3}}\frac{1}{\kappa ^{3/2}}\frac{%
\Gamma \left[ \kappa +1\right] }{\Gamma \left[ \kappa -1/2\right] }
\label{13}
\end{equation}

For a variation of the spectral index kappa Figure 1 demonstrates the different behavior
between the one dimensional case of the distribution function (12), generally used for
space applications and modeling, as compared to equation (6) resulting from the modified
entropy approach. Since (6) and (12) differ only in the exponent no exact mapping is possible
when substituting $\kappa \rightarrow \kappa+1$. Nevertheless, highly similar structures
are found for both with the appropriate values of kappa, hence favoring relation (6)
for astrophysical velocity space distribution modeling due to the physical background.

\begin{figure}[htb]
    \centering
    \includegraphics[height=3in]{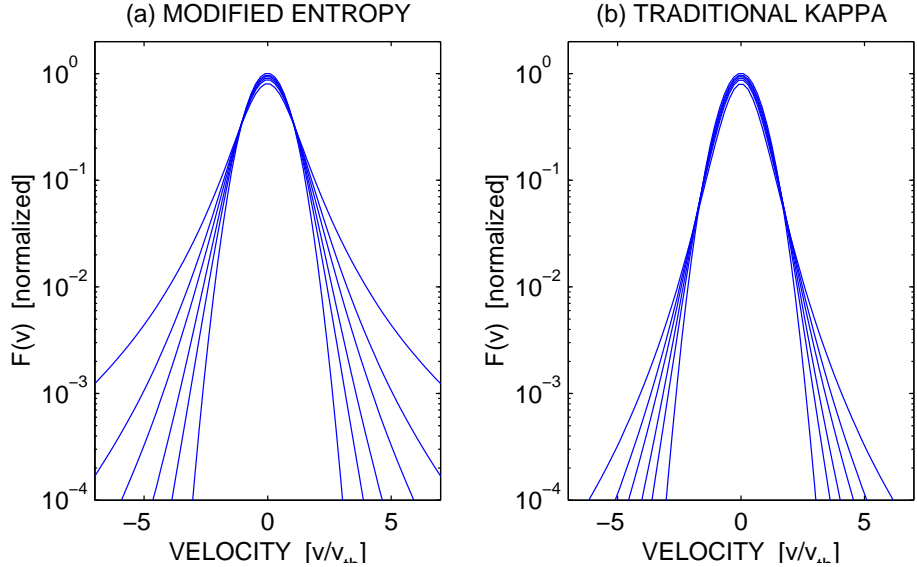}
    \caption{Schematic plot of the family of kappa-distributions.
The subplots (a) and (b) demonstrate the one-dimensional functional
dependence of equation (6) and (12), respectively. For both,  $\kappa =2$\
corresponds to the outermost curve followed by values $\kappa =3,4,6$ and $10
$. The innermost curve represents with $\kappa =\infty $ an isotropic
Maxwellian.}
    \label{Fig.1}
\end{figure}

An effective thermal speed $\theta =v_{th}\sqrt{(\kappa -3/2)/\kappa }$ is commonly
defind from the moments of the distribution function (12) where the
spectral index kappa is conventionally limited to positive values $\kappa >3/2$
and we note that negative values of kappa have never been
considered for space plasma applications. Contrary, the generalization to negative
values of kappa identical to $q\geq 1$, discussed in the framework of nonextensive
statistics as well, generates according to equation (\ref{6}) or (\ref{10}) a thermal
cutoff at the maximum allowed velocity

\begin{equation}
v_{\max }=\sqrt{\kappa \frac{2k_{B}T}{m}}  \label{14}
\end{equation}

hence providing an additional physical interpretation of the spectral index kappa
of non-thermal plasma structures. To our knowledge there is presently no
observational evidence for distributions corresponding to $q\geq 1$ in space
plasmas available. On the other hand it should be mentioned that any observed
thermal cutoff in velocity distributions would not have been recognized or
interpreted as being a consequence of a possible nonextensivity of the system.

Finally we note that Collier \cite{Collier95} has evaluated the Boltzmann entropy for
kappa-distribution functions. Upon introducing the corresponding nonextensive
velocity space distribution the resulting kappa-dependent entropy is found to
obey qualitatively the same functional dependence. Since the suprathermal tails
are more pronounced in the modified entropy case as compared to the standard
kappa function, see Figure 1, also the entropy enhancement for low values of
the spectral index, corresponding to a large fraction of suprathermal particle
populations, is increased relatively to the traditional kappa-distribution
case. Both approaches merge for large values of kappa towards the Maxwellian limit.

\section{Conclusions}

Nonextensive statistics was successfully applied to a number of
astrophysical and cosmological scenarios. Those include stellar polytropes
\cite{Plastino93}, the solar neutrino problem \cite{Kaniadakis96},
peculiar velocity distributions of galaxies \cite{Lavagno98} and generally
systems with long range interactions and fractal like space-times.
Cosmological implications were discussed \cite{Torres97} and
recently an analysis of plasma oscillations in a collisionless
thermal plasma was provided from q-statistics \cite{Lima00}.
On the other hand, kappa-distributions are highly favored in any kind of space plasma
modeling \cite{Mendis94} among others, where a reasonable physical
background was not apparent. A comprehensive discussion of kappa distributions in view of
experimentally favored non-thermal tail formations is provided by Leubner and
Schupfer \cite{Leubner00b} where also typical values of the index $\kappa $ are quoted and referenced
for different space plasma environments.

In the present analysis the missing link
to fundamental physics is provided within the framework of an entropy modification
consistent with nonextensive statistics. The family of kappa distributions are obtained
from the positive definite part $\frac{1}{2}\leq \kappa \leq \infty$,
corresponding to $-1\leq q\leq 1$ of the general
statistical formalism where in analogy the spectral index kappa is a measure
of the degree of nonextensivity. Since the main theorems of the standard
Maxwell-Boltzmann statistics admit profound generalizations within
nonextensive statistics \cite{Plastino94,Rajagopal95,Rajagopal96,
Chame97,Lenzi98}, a justification for the use of
kappa-distributions in astrophysical plasma modeling is provided from fundamental physics.


\end{document}